# Enhancing the Google imagery using a wavelet filter

**Amelia Carolina Sparavigna**
Dipartimento di Fisica, Politecnico di Torino
C.so Duca degli Abruzzi 24, Torino, Italy

**Abstract**
In some previous papers we proposed the use of free software for a processing of the Google satellite imagery. Here we discuss the use of a wavelet filter for the same purposes. This filter is a tool included in a freely downloadable software (Iris), well-known for the processing of astronomical images. Combining the image obtained after applying the wavelet filter, with an image created with Gimp and AstroFracTool, the visibility of the landforms, as obtained from Google Maps, is strongly increased. Among several possible examples, we proposed a crater, a paleochannel and the Great Bend of the Nile.

Google Maps is a web service providing high-resolution satellite images. Besides covering many urban areas, it gives a high-resolution imagery of all the Egypt's Nile Valley and many other regions in Africa, being then a quite powerful tool of "organizing the world's information geographically", as told by Lars Rasmussen, one of its creators [1].
Using Google Maps, we proposed in Ref.2 a survey of the Bayuda region in Sudan, that is, the region surrounded by the Great Bend of the river Nile, enhancing the images of the maps by processing them with two freely available tools. One is the tool we called AstroFracTool, based on fractional calculus, able to mark the edges without loosing the whole image texture [3]. An adjustment of contrast and brightness of the image obtained after applying AstroFracTool is achieved by means of Gimp, that is, the GNU Image Manipulation Program. Two following papers [4,5], demonstrated that this processing, applied to the Google Maps (or ACME Maps), is able to show the details of craters, large or very small in diameter too.
Here, we propose the use of an other freely downloadable software. It is Iris, by Christian Buil, a well-known tool for processing astronomical images [6]. Iris has many pre-processing and processing filters: among them we find the wavelet filter. Launching the corresponding mask, we see five adjustable resolution levels. Changing interactively the levels, it is possible to enhance all the features, small or large, of the image texture.
To see how the filter is working, let us look at Fig.1. Panel 1a contains the original image from Google of a rocky region in the Bayuda desert. The panel 1b shows the same area after processing with the wavelet filter, increasing only the finest and the fine levels, and fixing at very small values the medium, large and largest resolution levels. Note that we clearly see all the minute details of the texture. In panel 1c, there is instead the result obtained increasing the large and largest levels and fixing the medium, fine and finest to very low values. In this case, we see only the large features of the texture. Finally, adjusting all the levels to the values reported by the snapshot of the Iris wavelet filter, we have image 1d.
Actually, this example and those reported in [2,4,5] are demonstrating that it is possible to improve the imagery of landforms applying the three free software tools, Gimp,

AstroFracTool and Iris in several manners.
One possible way is the following, among the many possibilities at disposal:

a) process with AstroFracTool the original image for enhancing  edges and then adjust the brightness and contrast of the resulting image with Gimp, so having an Output A,
b) process the original image with the wavelet filter of Iris, obtaining an Output B,
c) and, finally, prepare a proper combination of output A and output B,  with the Gimp merging tool.

Let us show the results we can see for three cases studies: a crater near the Nile, a paleochannel of the river, previously observed in a space Shuttle mission [7], and the Bayuda Great Bend.

*A crater near the Nile*: The Nile has two major tributaries, the White Nile and Blue Nile, coming from the rainfall regions in equatorial Africa and Ethiopia. Flowing through the Sahara Desert, the course of the river is mainly controlled by the bedrock fabrics.  In northern Sudan, the Nile forms the great bend of the Bayuda desert. Besides the Bayuda great bend, the River Nile has several bends where it is changing its flowing direction. One bend is at 19.937N,30.296E. Before the bend, from the river departs a small branch, which creates the Arduan island. This island is known for Neolithic settlements. Few kilometres N-W from this island, we see a crater-like landform, with a diameter of 2.5 km (the author is not able to give references on this crater). The Google terrain clearly shows the rim of the crater.

The previously proposed procedure is applied to a larger scale area about the crater, as in Fig.2a. Output A is reproduced in panel 2b. A proper choice of the wavelet levels gives Output B in panel 2c. Combining the two outputs, we have the final image 2d. Note that the crater has a form completely different from the texture of the background. For this region of the Nile, Google Maps has the possibility to reach a very high resolution of satellite imagery. Zooming,  image in Fig.3 is obtained: it is the merging of two outputs A and B. Note the rim of the structure and the dry drainage network of the area. Fig.2 and 3 is processing a large region, but the procedure can repeat for the crater only with very high resolution.

*A paleochannel*: The great bend of the Bayuda desert was studied with remote sensing imagery, acquired with the SIR-C/X-SAR imaging  system, during two flights of the NASA space shuttle Endeavor in 1994 [7]. These data reveal how bedrock structures of different age control much of the Nile's course. During the mission, an ancient  channel of the Nile buried under the sand. The radar images were processed at NASA's Jet Propulsion Laboratory, Pasadena, and the University of Texas at Dallas. The discovery of the dry river channel, defined as a paleochannel, shows that the Nile was forced to abandon its bed and take up a new course to the south. This means that the region (Bayuda) has been tectonically active. In Fig.3.of Ref.7, this paleochannel is clearly shown.

Let us check whether Google Maps is able to shown this paleochannel or not. Surprisingly, we find it with very high details after processing. In Fig.4, panel a, original Google image, the paleochannel is faintly detectable. Panel 4b reports Output A, whereas panel 4c the Output B obtained with Iris. The merging is proposed in Fig 4d. The old river is completely evidenced, with its drainage basin.

*The Great Bend*: Figure 4 shows a small part of the Great Bend of the Nile river, during its flow through the Sudanese desert. This huge bend, from the fourth to the sixth cataract, of the river surrounds the desert of Bayuda. It is a very dry land, characterised from the basaltic rocks of ancient volcanoes. A cluster of well-preserved volcanoes are present in the northern part of Bayuda Desert. These volcanoes are creating a more or less continuous field, with isolated centres of eruption. In Ref.8, the authors have observed the subsidence of surface or near-surface of the region. This region was then subjected to an activity able to change the flow of the river and related channels.

Let us see a part of the Great Bend, with the Google Maps. Fig.5 reproduces the image as resulting from combining the two outputs A and B. This image demonstrated that this area is extremely rich for further investigations on the past of the Nile. Researches can have then the possibility to observe all the dry basin of river, and repeat or propose new studies on the history and past behaviour of such an important river.

As a conclusion, in this paper we have shown that the good processing results, obtained with AstroFracTool and Gimp, can be furthermore improved with the combined action of wavelet filtering. Therefore, with the suitable image processing, it is possible to obtain quite interesting geographically information from the Google Maps, at least for the desert landforms. Other areas are under investigations, to find a suitable method.

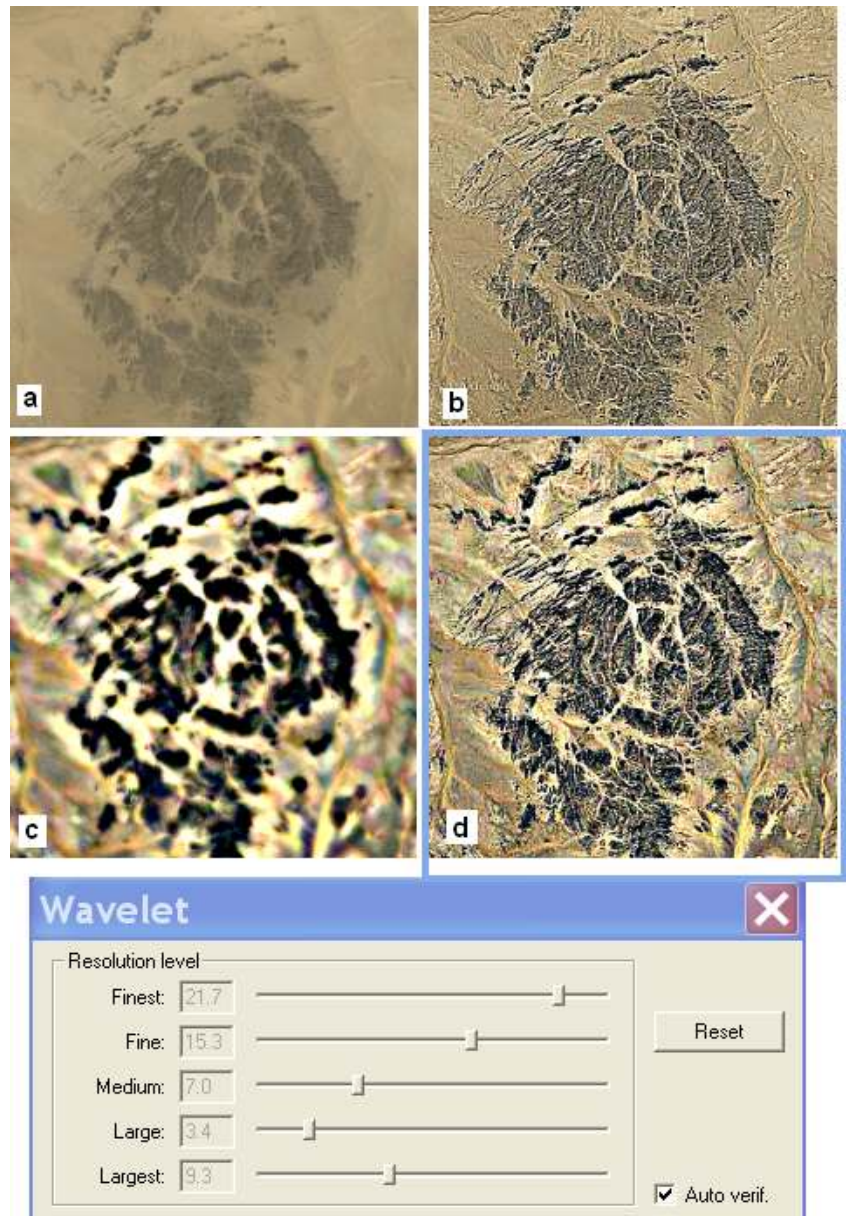

Figure1. The figure shows how the wavelet filter is working. Panel 1a contains the original image from Google of a rocky region in the Bayuda desert. 1b shows the same area after processing with the wavelet filter, increasing the finest and fine levels. Note the many minute details. Panel 1c shows the result of increasing the large and largest levels only. Adjusting all the levels, it is possible to have 1d. The snapshot of the corresponding values of filtering is placed in the lower part of the figure.

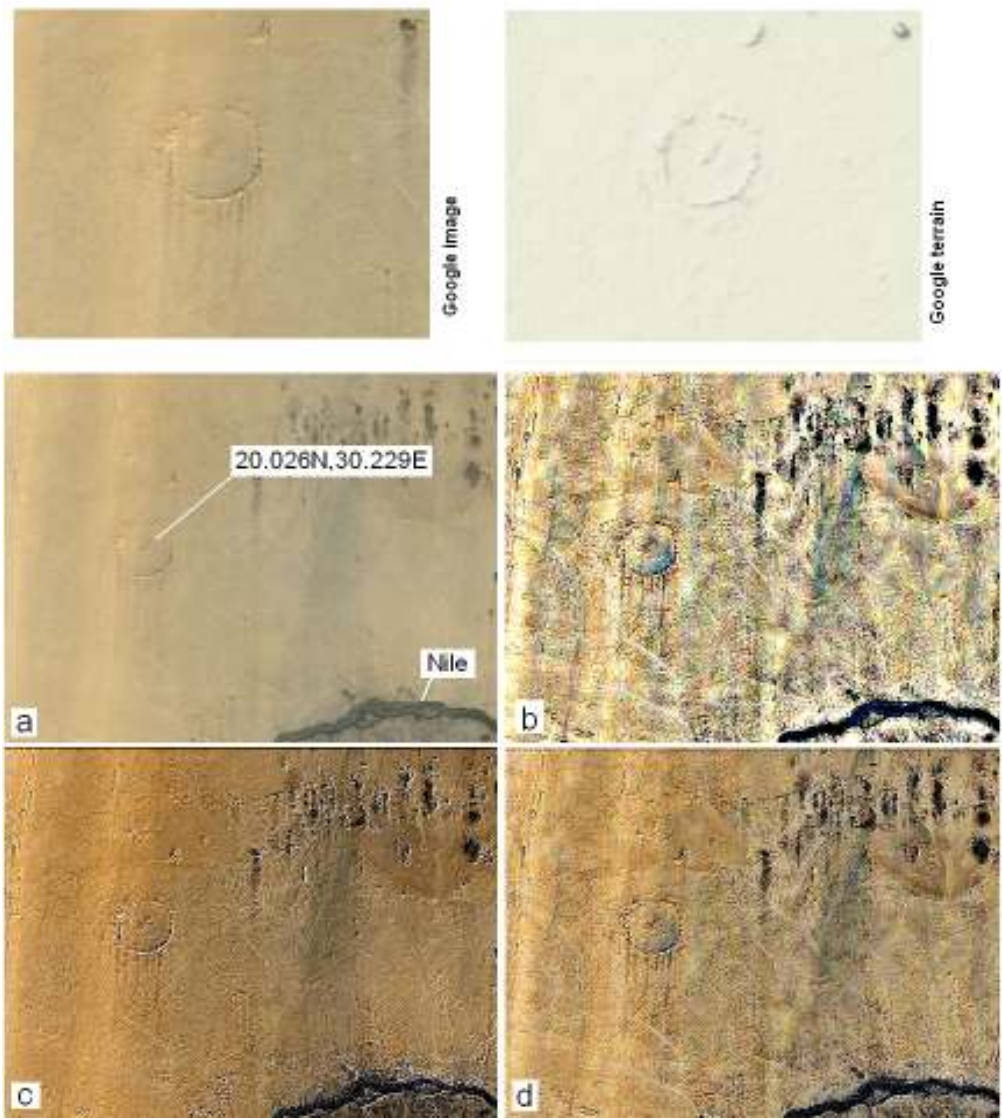

Figure2: A crater near the Nile. The River has several bends where it is changing its flowing direction. One bend is at 19.937N,30.296E. Before the bend, from the river departs a small branch, which creates the Arduan island. Few kilometres N-W from this island, there is a crater-like landform, with a diameter of 2.5 km. In the upper panels, the original Google image and the corresponding Google terrain. According to the proposed procedure (see text for explanation) Output A (panel 2b) and Output B (panel 2c) have been prepared. Combining the two outputs, we have the final image 2d. Note the form of the crater superimposed to the background fabric.

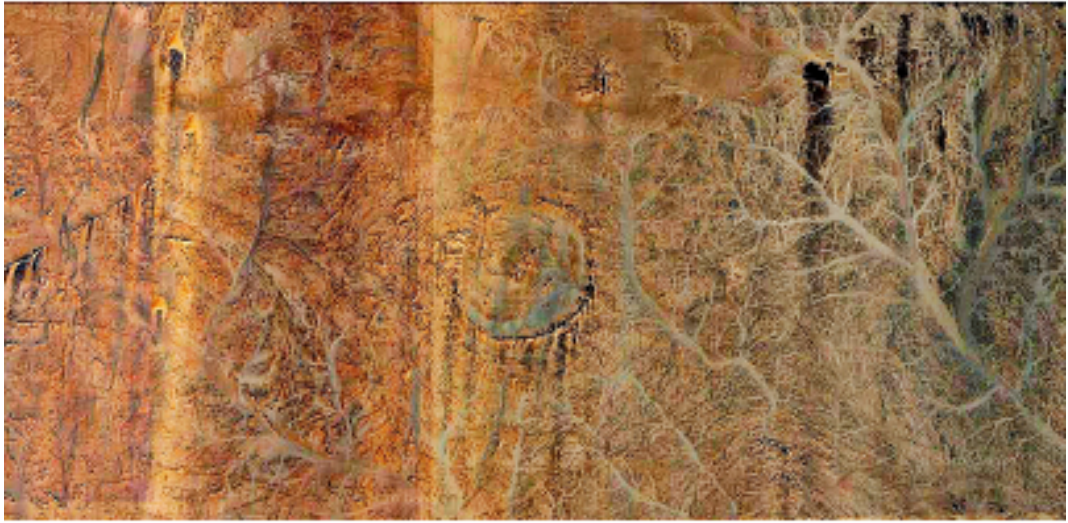

Figure 3: A higher resolution image of the crater, as obtained from merging of outputs A and B of the processing. Note the rim of the structure and the dry drainage network of the area.

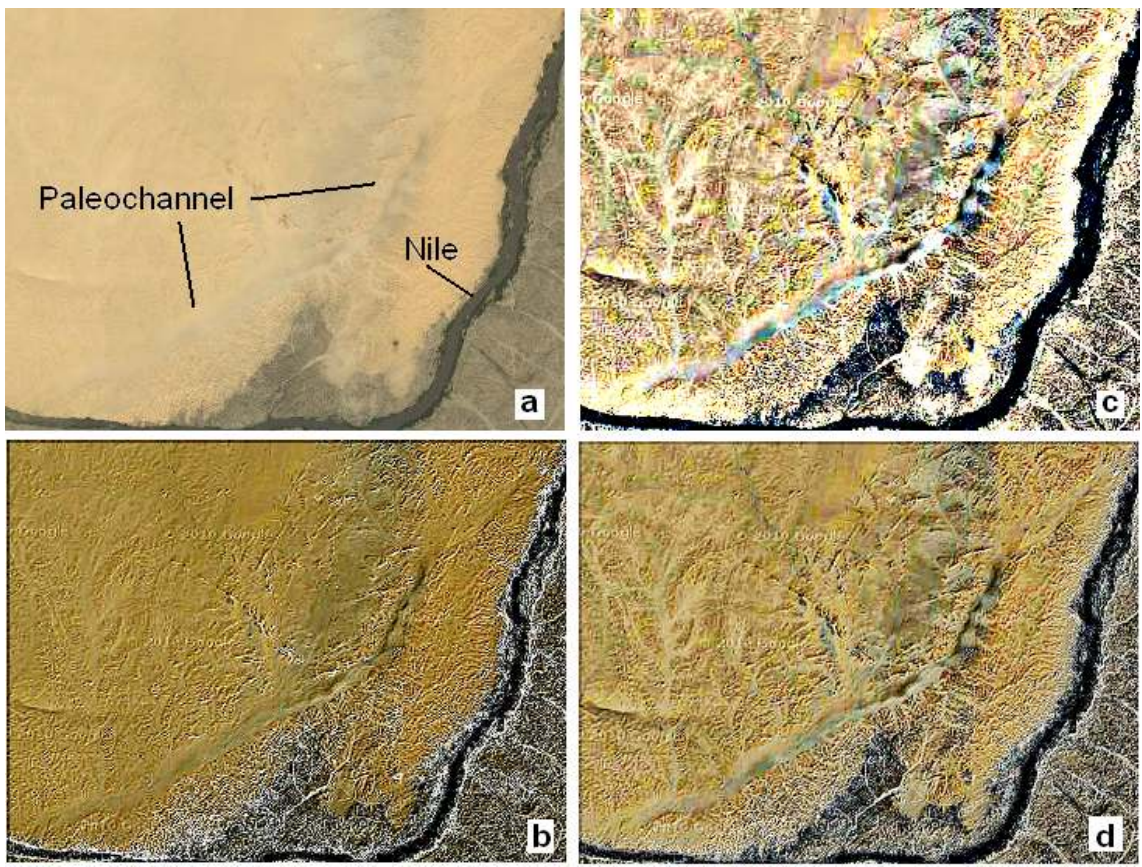

Figure 4: A paleochannel of the Nile. Panel 4a shows the original Google image. Panel 4b reports Output A, whereas panel 4c the Output B obtained with Iris. The merging is proposed in 4d. The old channel of the river is completely evidenced, with its drainage basin.

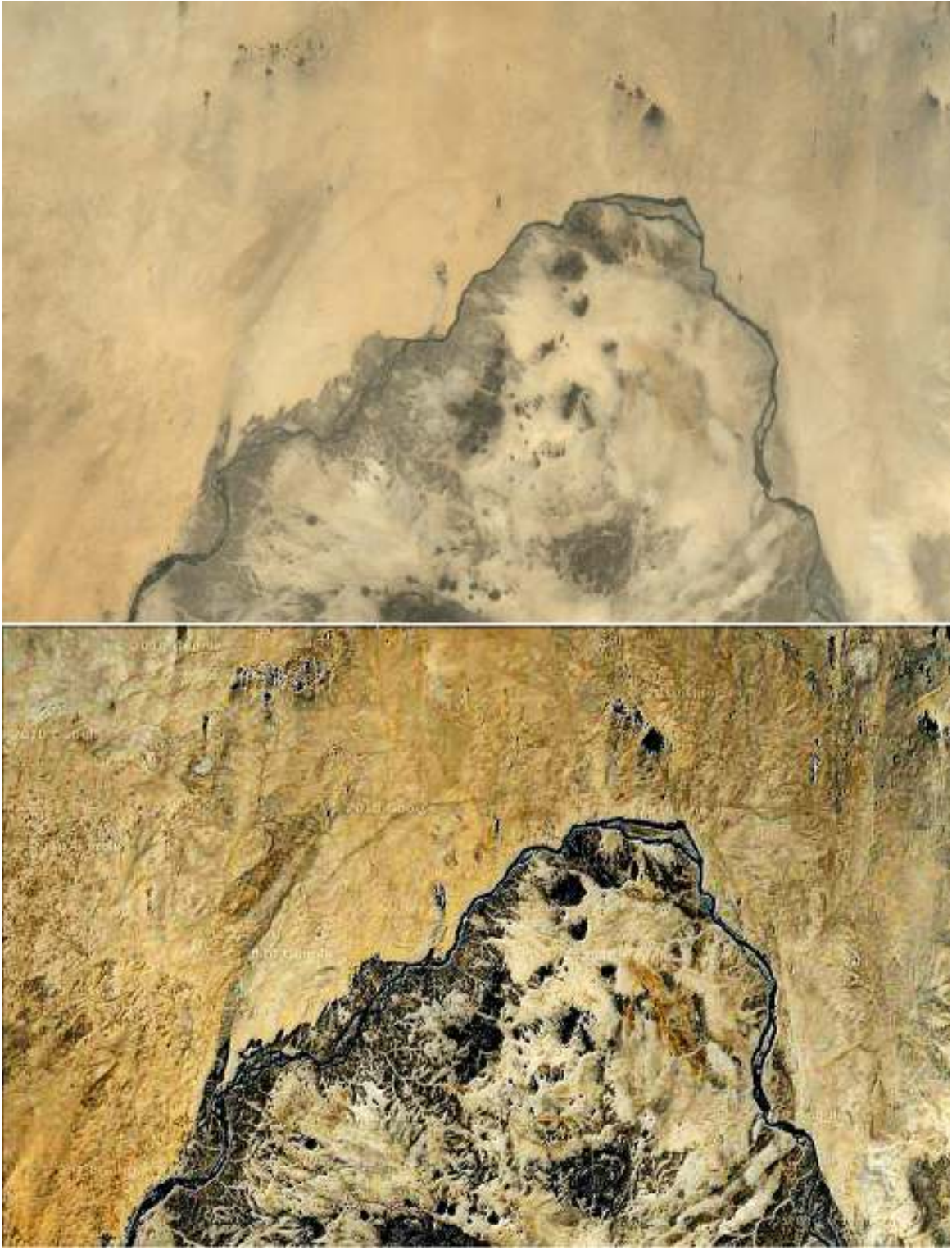

Figure 5: The Great Bend of Bayuda, original Google image (up) and after processing (down). We have the possibility to observe the complexity of all the networks of dry basin of the river.